\documentstyle[12pt]{article}
\title{The Fourth Root of Gravity\thanks{Submitted to the 1993
competition of the Gravity Research Foundation.}}
\author{Kevin Cahill\\
Department of Physics and Astronomy\\
University of New Mexico\\
Albuquerque, New Mexico 87131-1156\\
cahill@unmb.bitnet}

\begin{document}
\maketitle
\begin{abstract}
By attaching basis vectors to the components
of matter fields, one may render
free action densities
fully covariant.
Both the connection and the tetrads
are quadratic forms in these basis vectors.
The metric of spacetime, which
is quadratic in the tetrads, is then
quartic in the basis vectors.
\end{abstract}
\section*{Introduction}
We are accustomed to writing matter fields
as column vectors with components $\psi_a$.
For a two-component right-handed spinor R,
we might write
\begin{equation}
R(x)=\pmatrix{R_1(x)\cr R_2(x)\cr}
\end{equation}
in a kind of mathematical baby talk.
What we should write is
\begin{equation}
R(x)=R_a(x) e^a(x)
\end{equation}
in which the complex basis vectors $e^a(x)$
for $a=1,2$ depend upon the spacetime point $x$.
It is not obvious how many components $e^a_r(x)$
these vectors ought to possess; in this essay
they will have two complex components.
We shall see that by attaching these vectors $e^a(x)$
to the components $R_a(x)$, we may render
the free Dirac action density fully covariant.
\par
\section*{The Effect of the Vectors $e^a(x)$}
Let us consider the free action density $S$
for the spinor $R$,
\begin{equation}
S = i R(x)^\dagger \sigma^\mu \partial_\mu R(x),
\end{equation}
where $\sigma^\mu = (1, \vec \sigma)$ are the
Pauli matrices.
If we write $R$ with its vectors $e^a$,
then we find
\begin{equation}
S = i R_c(x)^\dagger e^{c \dagger}_r(x)
\sigma^\mu_{rs} \left[e^a_s(x) R_{a,\mu}(x)
+ R_a(x) e^a_{s,\mu}(x)\right].
\end{equation}
By inserting the two-by-two identity matrix
\begin{equation}
1 = e^a(x) e_a(x)^\dagger
\end{equation}
before the term $R_a(x) e^a_{s,\mu}(x)$
and by renaming some indices, we may write
the action density in the form
\begin{equation}
S = i R_c(x)^\dagger e^{c \dagger}(x)
\sigma^\mu \left[e^a(x) R_{a,\mu}(x)
+ R_b(x) e^a(x) e_a(x)^\dagger \cdot e^b_{,\mu}(x)\right].
\end{equation}
Thus if we identify the connection $A_\mu(x)$ as
\begin{equation}
A_{\mu a}^b(x) = e_a(x)^\dagger \cdot e^b_{,\mu}(x)
\end{equation}
and the tetrad $V^\mu_i(x)$ by the relation
\begin{equation}
V^\mu_i(x)\sigma^{i c a} = e^{c \dagger}_k(x)
\sigma^\mu_{kl} e^a_l(x),
\end{equation}
or equivalently as the trace
\begin{equation}
V^\mu_i(x) = (1/2) {\rm tr} ( \sigma^i e^\dagger \sigma^\mu e ),
\end{equation}
then we find for the action density
the usual expression
\begin{equation}
S = i R_c(x)^\dagger V^\mu_i(x)\sigma^{i c a}
D_{\mu a}^b R_b(x)
\end{equation}
with
\begin{equation}
D_{\mu a}^b R_b(x) = \left[\delta_a^b \partial_\mu
+ A_{\mu a}^b(x) \right] R_b(x).
\end{equation}
\par
A similar massage performed upon
the action density
\begin{equation}
iL^\dagger \sigma_\mu \partial_\mu L
\end{equation}
of the left-handed spinor $L(x) = L_a(x) f_a(x)$
yields a left-handed tetrad $W^\nu_j(x)$.
In terms of these tetrads
the metric of spacetime is~\cite{kc}
\begin{equation}
g^{\mu \nu} (x) = (1/2) ( V^\mu_i W^\nu_i + V^\nu_i W^\mu_i).
\end{equation}
\par
Since the metric is quartic in the vectors
$e^a$ and $f^b$,
these basis vectors may be thought of as
fourth roots of the metric.
The present formalism expresses both the metric $g_{\mu\nu}$
and the connection $A^b_{\mu a}$ in terms of the vectors
$e^a$ and $f^b$, which in turn are parts of
the vectors $R$ and $L$.
\par
\section*{A Minimal Theory of Gravitation}
As the action for the metric $g_{\mu\nu}$ itself,
one may choose among various actions constructed from
the vectors $e^a$ and $f^b$ and from
the connection $A^b_{\mu a}$.
The simplest and most conservative choice
is to use metric $g_{\mu\nu}$
to build the usual Einstein action density
\begin{equation}
S_g = \sqrt{-g(x)} \, R(x)
\end{equation}
and to require the matrices $e$ and $f$ to
have determinant unity and to
obey the relation
\begin{equation}
e = f^{-1 \dagger}.
\end{equation}
The two tetrads $V$ and $W$ are then identical,
and the gravitational field is described
by three complex numbers at each point $x$
of spacetime.
\section*{Acknowledgements}
I should like to thank Srikanth Raghavan
and Mark Frautschi for helping me.
This work was supported in large part
by the U.~S. Department of Energy
under contract DE-FG04-84ER40166.

\end{document}